\documentclass[fleqn,12pt,twoside]{article}
\usepackage{amsmath,amssymb,citesort,enumerate}
\usepackage{espcrc1}
\usepackage{epsfig}

\newcommand{\eqeqref}[1]{Eq.~\eqref{#1}}
\newcommand{\eqseqref}[1]{Eqs.~\eqref{#1}}
\newcommand{\refref}[1]{Ref.~\cite{#1}}
\newcommand{\refsref}[1]{Refs.~\cite{#1}}

\newcommand{\figref}[1]{Fig.~\ref{#1}}

\newcommand{\beq}{\begin{equation}}
\newcommand{\eeq}{\end{equation}}
\newcommand{\bea}{\begin{eqnarray}}
\newcommand{\beas}{\begin{eqnarray*}}
\newcommand{\beau}[1]{\begin{equation} \label{#1} \begin{array}{rcl}}
\newcommand{\eea}{\end{eqnarray}}
\newcommand{\eeas}{\end{eqnarray*}}
\newcommand{\eeau}{\end{array} \end{equation}}
\newcommand{\bay}{\begin{array}}
\newcommand{\eay}{\end{array}}
\newcommand{\bals}{\begin{align*}}
\newcommand{\eals}{\end{align*}}

\newcommand{\ds}{\displaystyle}

\newcommand{\ra}{{\rightarrow}}

\newcommand{\vev}[1]{\langle #1 \rangle}

\title{Nuclear modifications of fragmentation functions 
	and rescaling models
	\thanks{Talk presented at nthe ``European workshop on the QCD
	structure of the nucleon'' (QCD-N'02), Ferrara (ITA), April
	3rd-6th, 2002.}
	\thanks{This work is (partially) funded by the European
	Commission IHP program under contract HPRN-CT-2000-00130.}}

\author{Alberto Accardi
	\address[HDU]{Institut f\"ur Theoretische Physik der
	Universit\"at  Heidelberg, Germany}\thanks{E-mail address: accardi@tphys.uni-heidelberg.de}
        and
        Hans J. Pirner
	\addressmark[HDU]\address{Max-Planck-Institut
	f\"ur Kernphysik, Heidelberg, Germany}\thanks{E-mail address: pir@tphys.uni-heidelberg.de}
}

\begin{document}

\maketitle

\begin{abstract}
We discuss nuclear modification of fragmentation functions in the
context of the so-called ``rescaling models''. These models implement
partial deconfinement inside nuclei by modifying the fragmentation
functions perturbatively. We apply these models to the analysis of
nuclear hadron production in deep inelastic scattering processes at
the HERMES and EMC experiments. \\
\end{abstract}



\begin{figure}[t]
\begin{center}
\begin{minipage}[t]{11cm}
\begin{center}
\parbox{5.3cm}{\vskip-.5cm
\epsfig{figure=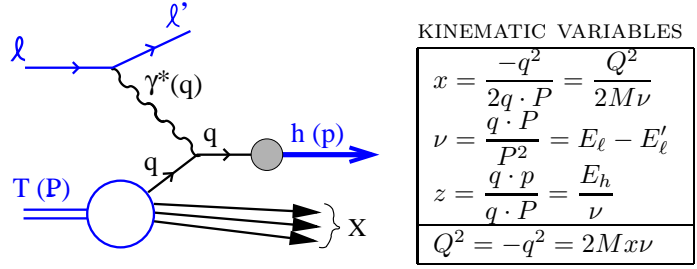,width=5cm}}
\parbox{3.8cm}{
{\footnotesize \sc kinematic variables} \vspace*{.1cm}\\
\begin{tabular}{|l|} 
\hline 
\footnotesize 
$\ds x = \frac{-q^2}{2q\cdot P} = \frac{Q^2}{2M\nu}$ \vspace*{0cm}\\
\footnotesize 
$\ds \nu = \frac{q\cdot P}{P^2} = E_\ell -E_\ell ' $ \vspace*{0cm} \\
\footnotesize 
$\ds z = \frac{q \cdot p}{q\cdot P} = \frac{E_h}{\nu}$  \vspace*{0cm}\\
\hline
\footnotesize
$\ds Q^2 = -q^2 = 2 M x \nu $ \vspace*{0cm}\\ \hline
\end{tabular}}
\vskip-.5cm
\caption{\footnotesize Kinematic variables for DIS scattering; four-moments are indicated in
parentheses.}
\label{fig:DIS}
\end{center}
\end{minipage}
\end{center}
\vskip-.7cm
\end{figure}

In deep inelastic scattering a projectile lepton $\ell$ emits a virtual 
photon $\gamma^*$, which scatters on a quark $q$ from the target $T$, in
our case a deuteron $D$ or a heavier nucleus with atomic number $A$. The
struck quark fragments into the observed hadron $h$, see
\figref{fig:DIS}. We use kinematic variables as summarized in the same figure:
$x$ is the Bjorken's scaling variable, $\nu$ is the energy of the
virtual photon in the target rest frame, and $z$ is the
fraction of the energy transferred to the produced hadron. 

The experimental data on nuclear effects in hadron production 
are usually presented in terms of the {\it multiplicity ratio}
\begin{align} 
    R_M^h(z) = 
	\frac{\ds 1}{\ds N_A^\ell} \frac{\ds dN_A^h}{\ds dz}
	\bigg/ \frac{\ds 1}{\ds N_D^\ell} \frac{\ds dN_D^h}{\ds dz} \ ,
 \label{multratio}
\end{align}
where $N_A^\ell$ is the number of outgoing leptons in DIS processes on
a nuclear target of atomic number $A$ and $dN_A^h/dz$
is the $z$-distribution of produced hadrons in the same processes; the
subscript $D$ refers to the same quantities when the target is a
deuteron. A similar definition for the multiplicity ratio as a function
of $\nu$ may be written. 
In absence of nuclear effects the ratio $R^h_M$ would be equal to
1. The experimental observation that $R^h_M \neq 1$
\cite{SLAC,EMC,HERMES} has been explained theoretically in many ways: as
an effect of nuclear absorption of the produced hadrons
\cite{BC83,BG87}, in a
gluon-bremsstrahlung model for leading hadron production \cite{KNP96},
as an higher-twist effect \cite{hightwist}. In this short note we will
not discuss these models in detail (see \refref{Muccifora02}) and,
instead, will concentrate on the so-called {\it rescaling models}
\cite{NP84,CJRR84,DDD86}.  



\newpage

The starting point of rescaling models is to assume a change in the
confinement scale $\lambda_A$ in nuclei, compared to the confinement
scale $\lambda_0$ in free nucleons: 
\begin{align*}
	\lambda_A > \lambda_0 \ .
\end{align*}
This assumed partial deconfinement in nuclei affects both the
parton distribution functions (PDF) and the fragmentation functions
(FF). For what concerns PDF's \cite{NP84,CJRR84}, consider
a valence quark which carries a momentum $Q_0$ when it is 
confined on a scale $\lambda_0$. If the scale changes to  $\lambda_A$
it carries a corresponding momentum $Q_A$. Since there is no
other dimensionful scale, the product $Q\lambda$ must remain constant,
so that
\begin{align}
	Q_0 \lambda_0 = Q_A \lambda_A \ .
 \label{deconf}
\end{align}
Therefore, if we take $Q_0$ to be the initial scale for the DGLAP
evolution of distribution functions, we may set
\begin{align}
    q_f^A \big( x, Q_A=\frac{\ds \lambda_0}{\lambda_A} \, Q_0 \big) 
	= q_f\big(x,Q_0\big) \ ,
 \label{PDF0}
\end{align}
where $q_f$ is the distribution function of a quark of flavour $f$
in a free nucleon and $q_f^A$ is the same quantity 
when the nucleon is inside a nucleus. 
For FF's \cite{DDD86} a similar argument holds
because constituent quark and effective hadron masses are sensitive to
the confinement scale. Therefore if we take $Q_0$ to be the physical
threshold for hadron production we may set
\begin{align}
    D_f^{h|A} \big(x, Q_A=\frac{\ds \lambda_0}{\lambda_A} \, Q_0 \big) 
	 = D_f^h\big(x,Q_0\big) \ , 
 \label{FF0}
\end{align}
where $D_f^h$ is the fragmentation function of a quark of flavour $f$
into a hadron $h$ and $D_f^{h|A}$ is the nuclear modified
fragmentation function.
To extend \eqseqref{PDF0} and \eqref{FF0} to an arbitrary scale $Q$ we
apply pQCD evolution. The nuclear structure and fragmentation functions
evolve over larger interval in momentum compared to the corresponding
functions at the same scale Q, since the starting scale is smaller, see
\eqeqref{deconf}. The final result \cite{NP84,CJRR84,DDD86} is 
\begin{align}
	q^A_f(x,Q) & = q_f(x,\xi_A(Q) Q) 
 \label{rescPDF} \\
	D^{h|A}_f(z,Q) & = D^h_f(x,\xi_A(Q) Q) \ ,
 \label{rescFF}
\end{align}
where the {\it scale factor} $\xi_A(Q)$ is greater than one. These two
equations are the main tool which we will exploit.
There are two models for the computation of the scale factor:
\begin{enumerate}[a)]
\item \vskip0cm
the \underline{\it maximal deconfinement model} (NP) \cite{NP84}
assumes the onset of ``colour conductivity'' in nuclei and takes
$\lambda_A=R_A$, where $R_A$ is the nuclear radius, so that  
\[  
	\xi_A(Q) = R_A/R_p \ ;
\]
\item \vskip-.2cm
the \underline{\it partial deconfinement model} (CJRR) \cite{CJRR84} 
assumes the deconfinement scale $\lambda_A$ to be proportional to the
degree of overlap of the nucleons inside the given nucleus, and 
\[ 
	\xi_A(Q) = \big(\lambda_A/\lambda_0\big)
		^{\frac12 \frac{\alpha_s(Q_0)}{\alpha_s(Q)}} \ .
\]
\vskip-1cm \
\end{enumerate}
Note that the maximal deconfinement
model assumes a much larger scale factor, which results in a larger
nuclear modification of hadron production at high $z$. 

We calculate the multiplicity ratio \eqref{multratio}
 using the {\it rescaled} PDF \eqref{rescPDF}
and the {\it rescaled} FF \eqref{rescFF} in the leading order pQCD 
computation of $N_A^\ell$ and $dN_A^h/dz$:
\begin{align}
    \frac{1}{N_A^\ell}\frac{dN_A^h}{dz} & = \frac{1}{\sigma^{\gamma^*A}} 
	 \int_{\rm exp.\ cuts}\hspace*{-1.0cm}
	 dx\, d\nu \sum_f e^2_f\, q_f(x,\xi_A\,Q) \,
	 \frac{d\sigma^{\gamma^* q}}{dx\,d\nu} \, D_f^h(z,\xi_A\,Q)  
 \label{dnhdz} \\
    \sigma^{\gamma^*A} & = \int_{\rm exp.\ cuts}\hspace*{-1.0cm}
	 dx\, d\nu \sum_f e^2_f\, q_f(x,\xi_A\,Q) \,
	 \frac{d\sigma^{\gamma^* q}}{dx\,d\nu} \ , \nonumber
\end{align}
where $e_f$ is the electric charge of a quark of flavour $f$,  
$d\sigma^{\gamma^* q}/dx\,d\nu$ is the differential cross-section for
a $\gamma^* q$ scattering computed in pQCD at leading order, and
$Q^2=2Mx\nu$, with $M$ the nucleon mass. In the numerical computations
we used GRV98 parton distribution 
funtions \cite{GRV98} and Kretzer's parametrization of FF's at leading
order \cite{K}.

\begin{figure}[t]
\begin{center}
\begin{minipage}[t]{15.5cm}
\parbox{15.5cm}{\vskip-.5cm 
\epsfig{figure= 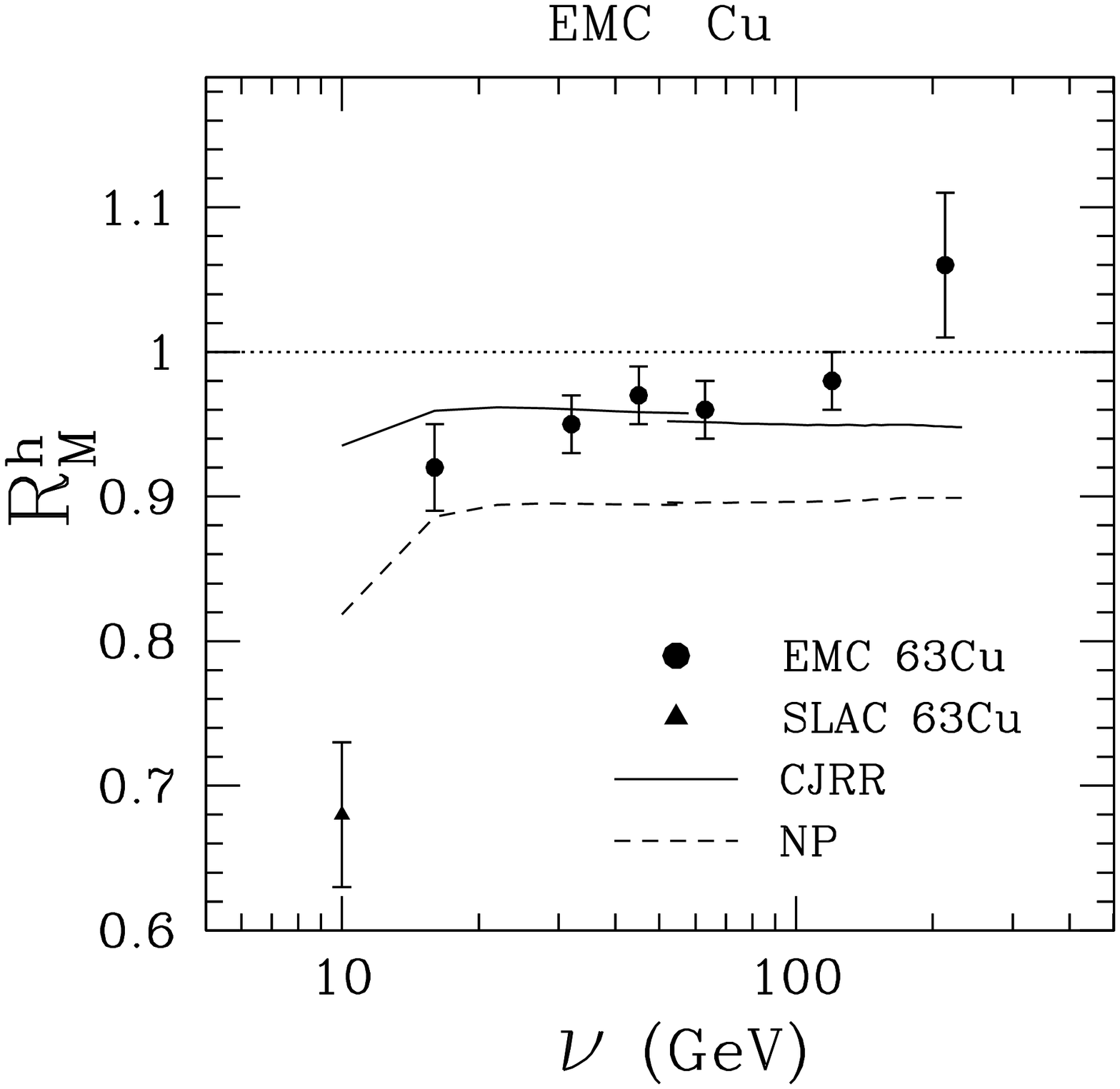,width=5cm}
\epsfig{figure= 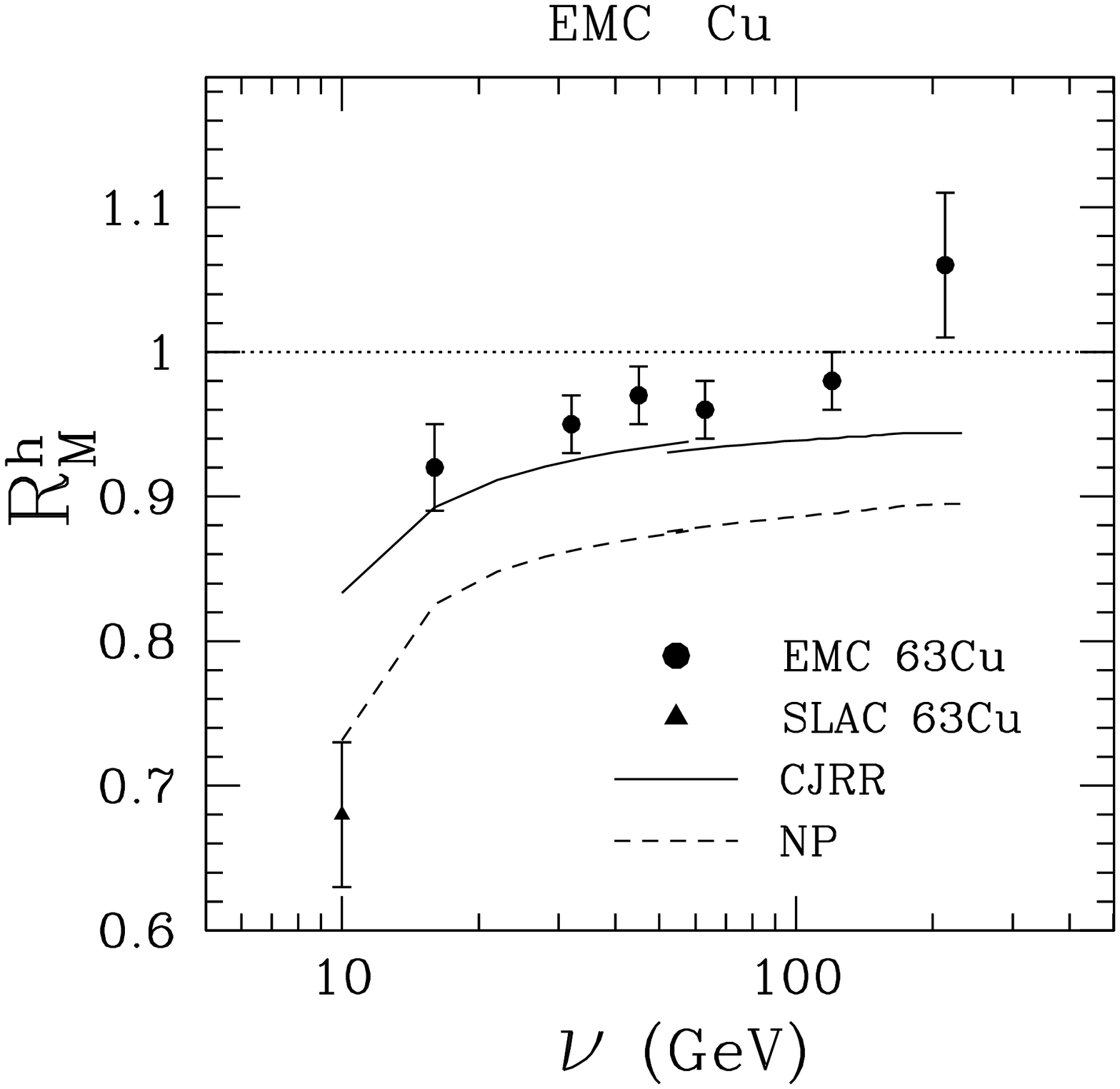,width=5cm}
\epsfig{figure= 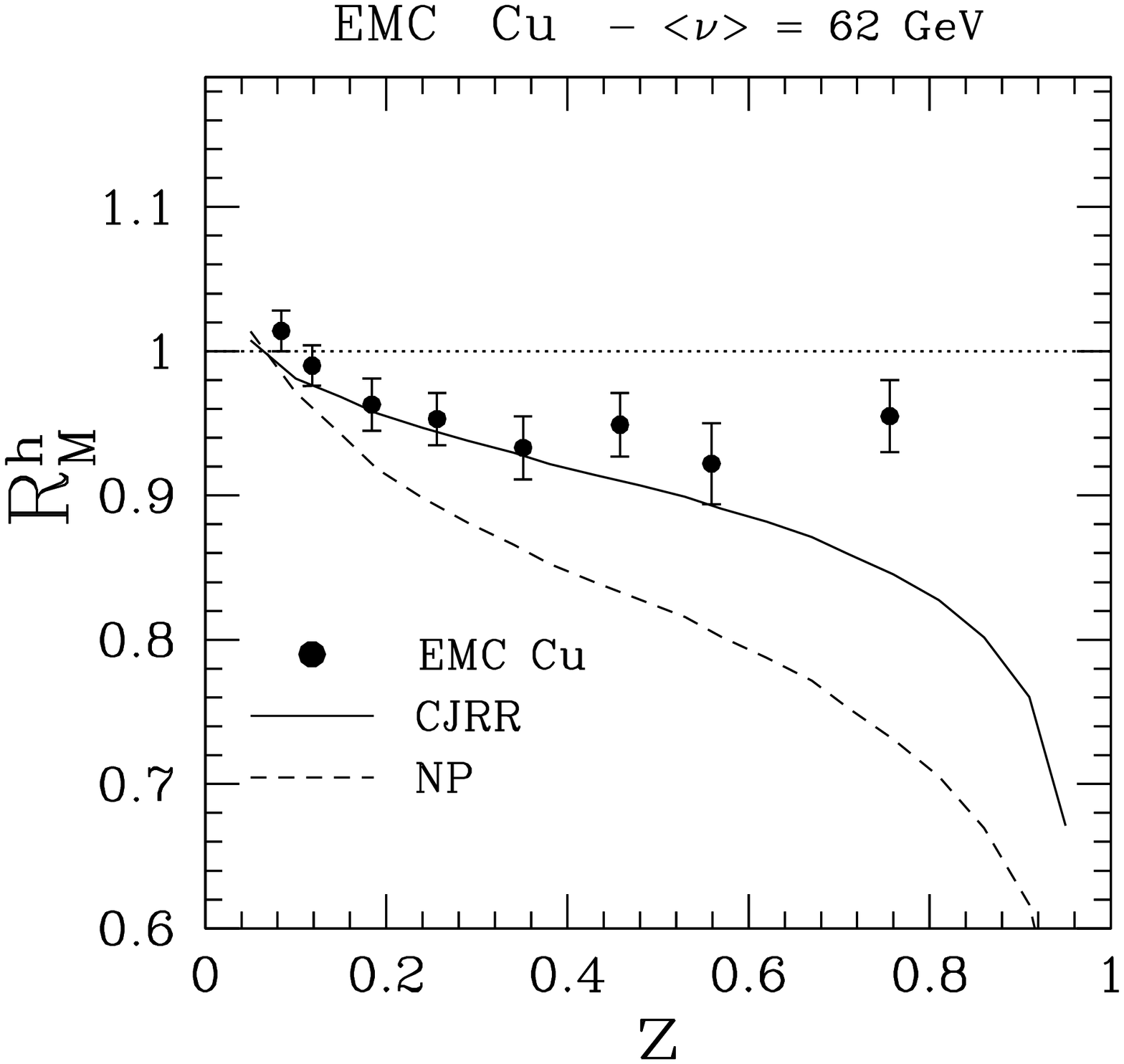,width=5cm} }
\vskip-.8cm
\caption{\footnotesize
Comparison of partial and maximal dconfinement models at 
EMC with a copper target. 
{\it a)} $\nu$-distribution with rescaling only and 
{\it b)} with rescaling and absorption;
{\it c)} $z$-distribution with rescaling and absorption. 
Data taken from \refref{EMC}.}
\label{fig:res1}
\end{minipage}
\end{center}
\vskip-.9cm
\end{figure}
  
The results for the $\nu$-dependence of the
multiplicity ratio for charged hadrons are shown in \figref{fig:res1}a and
compared to EMC data \cite{EMC}. The
maximal deconfinement model underestimates the data over nearly the whole
range. On the contrary the partial deconfinement model fit the data at
high $\nu$, but overestimates them at smaller $\nu$. Because of
Lorentz dilatation, at high $\nu$ the hadron is formed
mainly outside the nucleus and it is affected by rescaling effects
only. On the contrary, at small $\nu$ the hadron is formed inside the
nucleus and starts interacting with its nucleons, with some probability
of being absorbed.

To take into account nuclear absorption, we follow the analysis of
\refref{BC83,BG87}: the quark propagates in the nucleus with a
cross-section $\sigma_q$ for inelastic scattering on the
nucleons. Subsequently, it creates a pre-hadronic states, which has a
cross-section $\sigma_*$, and finally the observed hadron is formed,
which has a cross-section $\sigma_h$. Following the experimental
indications of EMC and HERMES, we take $\sigma_q=0$ and
$\sigma_*=\sigma_h=20$ mbarn for charged hadrons. With the latter
assumption we need to consider effectively only the {\it formation time}
$\tau_F$ for the pre-hadronic state \cite{BG87}:
\begin{align*}
    \tau_F = \left( \frac{-\ln(z^2) -1 + z^2}{1-z^2} \right)
	\frac{z\nu}{\kappa} \ ,
\end{align*}
where $\kappa$ is a parameter which we fix to $\kappa = 0.25$ GeV/fm.
Note that $\tau_F \ra \frac{1}{\kappa} (1-z)\nu$ as $z\ra1$,  
giving the formation time
suggested by the gluon bremsstrahlung model of \refref{KNP96}. 
Finally, nuclear absorption effects are
included in the computations by multiplying the integrand in
\eqref{dnhdz} by the {\it nuclear absorption factor} ${\cal N}_A$, which
represent the fraction of hadrons which escape from the nucleus:
\begin{align*}
    {\cal N}_A = \int d^2b \int_{-\infty}^{\infty} 
	dy \, \rho_A(b,y)\left[ S_A(b,y) \right]^{A-1} \ ,
\end{align*}
where $\rho_A(b,y)$ is the nuclear density normalized to 1 at transverse
and longitudinal coordinates $(b,y)$, and $S_A(b,y)$ is the survival
probability of a hadron produced in $(b,y)$:
\begin{align*}
    S_A(b,y) = 1 - \sigma_h \int_y^\infty \hspace*{-.2cm}dy\,'
	\rho_A(b,y\,')\left( 1 - e^{\,-(y\,'-y)/\tau_F} \right)  \ . 
\end{align*}
For the deuteron we used as a density the sum of the 
Reid's soft-core S- and D-wave functions squared \cite{ReidSC}. 
For heavier nuclei we used a Woods-Saxon density with radius
$R_A=1.12 A^{1/3} - 0.86 A^{-1/3}$ fm.

The $\nu$- and $z$-dependence of the multiplicity ratio at EMC and
HERMES after the inclusion of nuclear absorption are shown in
\figref{fig:res1}b and \figref{fig:res1}c. 
It is clearly seen that while the CJRR model gives
a nice description of the data, the NP model is ruled out.

\begin{figure}[t]
\begin{center}
\begin{minipage}[t]{15.5cm}
\parbox{15.5cm}{\vskip-.5cm 
\epsfig{figure=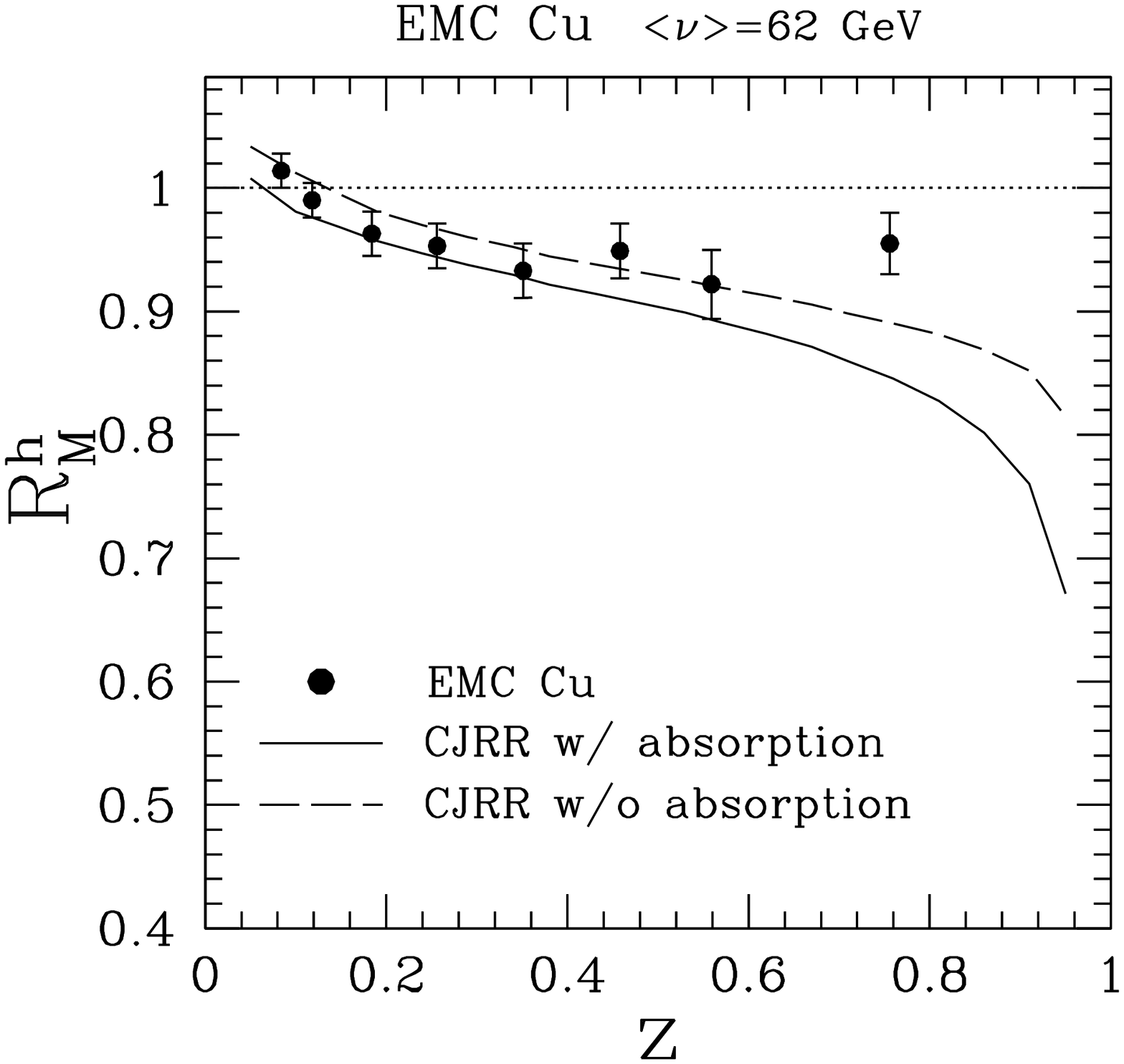,width=5cm}
\epsfig{figure=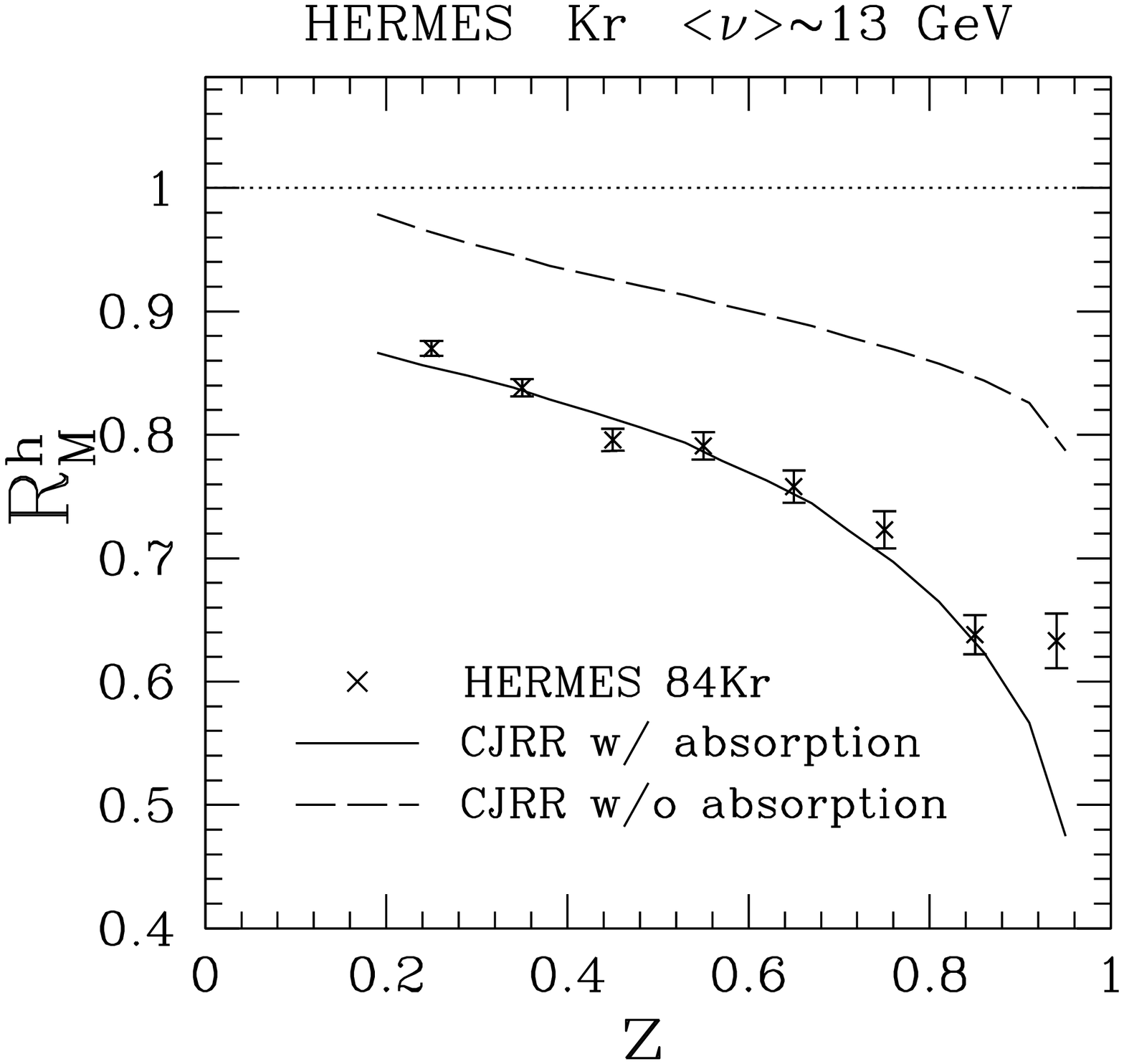,width=5cm}
\epsfig{figure=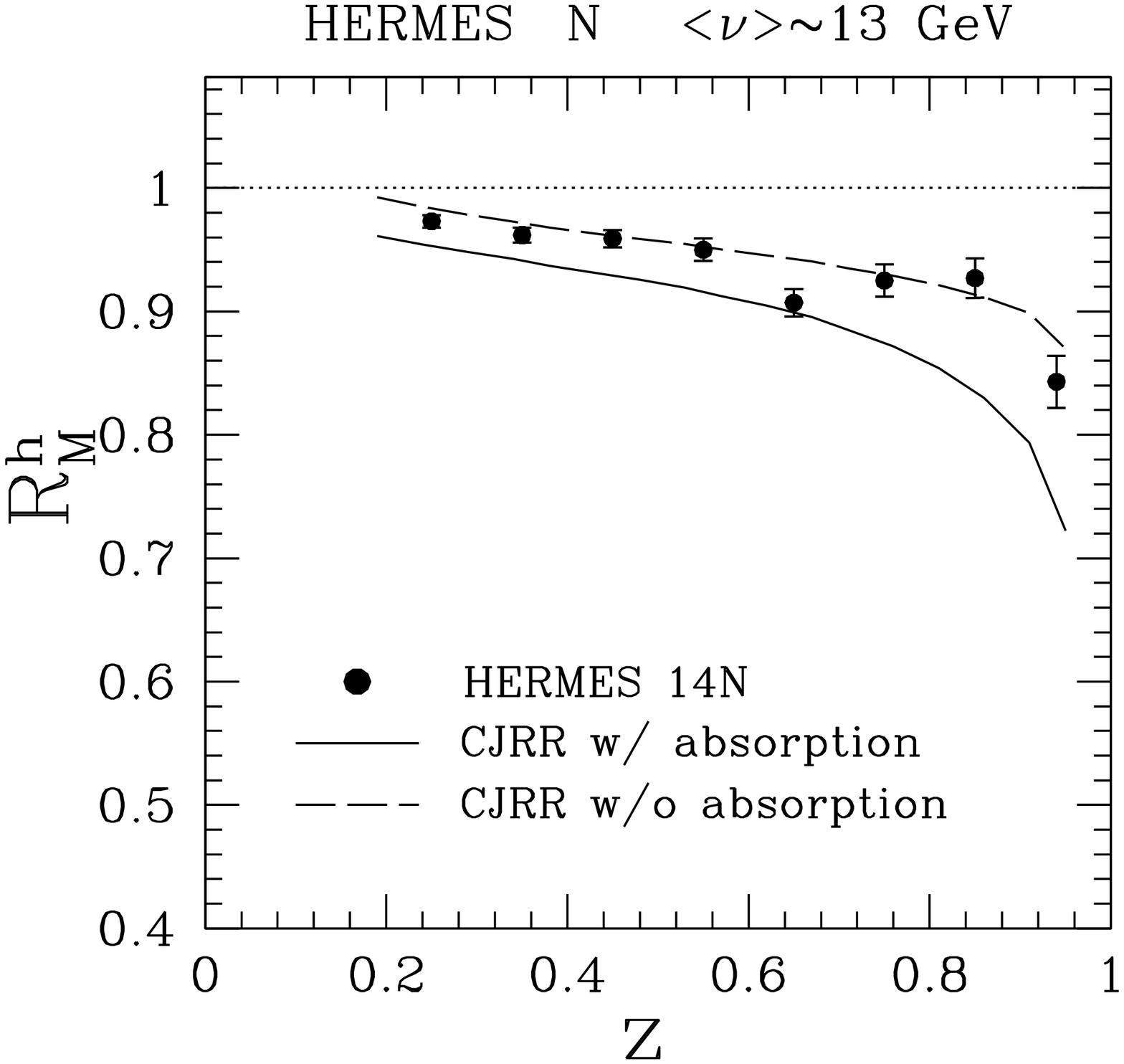,width=5cm} }
\vskip-.8cm
\caption{\footnotesize
Comparison of absorption effects in $z$-distributions with rescaling
only (dashed lines) and with 
rescaling and absorption (solid lines). From left to right: EMC with
copper target, HERMES with krypton and HERMES with nitrogen targets. 
Data taken from \refsref{EMC,HERMES}. Note that the scale in
\figref{fig:res2}a is different from \figref{fig:res1}c.}
\label{fig:res2}
\end{minipage}
\end{center}
\vskip-.9cm
\end{figure}

The average values of $\vev{\nu}$ available at HERMES are smaller than at EMC
and the absorption effects larger. In \figref{fig:res2} we
compare the $z$-distributions in the partial deconfinement models 
at EMC and HERMES before and after
inclusion of absorption. As expected, at EMC with a $^{63}$Cu target 
nuclear absorption is marginal and rescaling alone gives a
satisfactory description of the data. At HERMES with a $^{84}$Kr
target, which is comparable to copper, both effects are larger and
absorption is dominant, tending to mask rescaling effects. 
With a $^{14}$N target, which is smaller than krypton, both effects
are smaller. 
 


In summary, rescaling models (supplemented by nuclear absorption) 
are shown to be able to describe both HERMES and EMC data on the nuclear
modification of hadron production in DIS 
processes. While the maximal deconfinement model is ruled out by the
data - it assumes a too large deconfinement - the partial deconfinement
model is shown to be a good one. 
Further precise data at moderate and high $\nu$'s and for light and
heavy targets are needed to disentangle rescaling and formation time
effects. 

\vskip.3cm
{\bf Acknowledgments} 
\vskip.15cm

{ We are grateful to N.~Bianchi, P.~di~Nezza, B.~Kopeliovich and
V.~Muccifora for stimulating discussions and to R. Fabbri for 
technical support during the conference. }


\end{document}